\documentclass[aps,prb,showpacs,twocolumn,floats,superscriptaddress]
{revtex4}
\usepackage{graphicx}
\usepackage{bm}
\makeatletter
\def\lsim{\mathrel{\mathpalette\gl@align<}}
\def\gsim{\mathrel{\mathpalette\gl@align>}}
\def\gl@align#1#2{\lower.6ex\vbox
{\baselineskip\z@skip\lineskip\z@
\ialign{$\m@th#1\hfil##\hfil$\crcr#2\crcr\sim\crcr}}}

\makeatother

\begin{document}

\title{Slow quench dynamics of the Kitaev model: anisotropic critical point and effect of disorder}

\author{T. Hikichi}
\affiliation{Department of Physics and Mathematics, Aoyama Gakuin
University, Fuchinobe, Sagamihara 252-5258, Japan}

\author{S. Suzuki}
\affiliation{Department of Physics and Mathematics, Aoyama Gakuin
University, Fuchinobe, Sagamihara 252-5258, Japan}

\author{K. Sengupta}
\affiliation{Theoretical Physics Department, Indian Association for
the Cultivation of Science, Jadavpur, Kolkata-700032, India.}

\date{\today}

\begin{abstract}

We study the non-equilibrium slow dynamics for the Kitaev model both
in the presence and the absence of disorder. For the case without
disorder, we demonstrate, via an exact solution, that the model
provides an example of a system with an anisotropic critical point
and exhibits unusual scaling of defect density $n$ and residual
energy $Q$ for a slow linear quench. We provide a general expression
for the scaling of $n$ ($Q$) generated during a slow power-law
dynamics, characterized by a rate $\tau^{-1}$ and exponent $\alpha$,
from a gapped phase to an anisotropic quantum critical point in $d$
dimensions, for which the energy gap $\Delta_{\vec k} \sim k_i^z$
for $m$ momentum components ($i=1..m$) and $\sim k_i^{z'}$ for the
rest $d-m$ components ($i=m+1..d$) with $z\le z'$: $n \sim \tau^{-[m
+ (d-m)z/z']\nu \alpha/(z\nu \alpha +1)}$ ($Q \sim \tau^{-[(m+z)+
(d-m)z/z']\nu \alpha/(z\nu \alpha +1)}$). These general expressions
reproduce both the corresponding results for the Kitaev model as a
special case for $d=z'=2$ and $m=z=\nu=1$ and the well-known scaling
laws of $n$ and $Q$ for isotropic critical points for $z=z'$. We
also present an exact computation of all non-zero, independent,
multispin correlation functions of the Kitaev model for such a
quench and discuss their spatial dependence. For the disordered
Kitaev model, where the disorder is introduced via random choice of
the link variables $D_n$ in the model's Fermionic representation, we
find that $n \sim \tau^{-1/2}$ and $Q\sim \tau^{-1}$ ($Q\sim
\tau^{-1/2}$) for a slow linear quench ending in the gapless
(gapped) phase. We provide a qualitative explanation of such
scaling.

\end{abstract}


\maketitle

\section{Introduction}

Non-equilibrium dynamics of quantum systems near quantum critical
points has been a subject of intense study in recent years
\cite{pol1,dziar1}. During such dynamics, a quantum system passes
from one gapped phase to another via time evolution of a Hamiltonian
parameter $\lambda$ with a rate $\tau^{-1}$ and an exponent $\alpha$
($\lambda(t) = \lambda_0 |t/\tau|^{\alpha} {\rm Sgn}(t)$, where
${\rm Sgn}(x)=1(-1)$ for $x> (<) \,0$) through an intermediate
quantum critical point at $\lambda=0$. At the critical point, the
energy gap vanishes as $\Delta(\vec k) \sim |\vec k|^z$ where $z$ is
the dynamical critical exponent. Thus the dynamics becomes
non-adiabatic around a region near this point and the system fails
to remain at the instantaneous ground state leading to formation of
defects \cite{kibble1,zurek1,pol2,others1, sen1,pol3,sendutta}. The
density of these defects ($n$) and the residual energy produced in
the process ($Q$) scale with universal exponents: $n \sim \tau^{-\nu
d \alpha/(z \nu \alpha +1)}$ and $Q \sim \tau^{-(d+z)\nu
\alpha/(z\nu \alpha+1)}$, where $\nu$ is the correlation length
exponent and $d$ is the system dimension \cite{pol2,sen1}. It is
well-known that scaling laws do not change if the dynamics terminate
at the critical point \cite{pol3}. All of the above-mentioned
studies apply to isotropic critical points where the scaling of the
energy gap with the momentum is described by a single exponent $z$.
Recently, the anisotropic Dirac model with an anisotropic critical
point is studied and it was shown that one needs multiple exponents
to describe the scaling of the energy gap \cite{dutta1}. However
such studies have not been carried out in the context of the Kitaev
model and generic expressions for the scaling laws for $n$ and $Q$
for such critical points in arbitrary dimensions have not been
provided. Also, the effect of disorder on defect production in
models, where the Harris criterion allows for the existence of a
sharp quantum phase transition, has not been studied so
far\cite{sachdev1}.

In this work, we study several aspects of non-equilibrium slow
dynamics in the vicinity of both anisotropic critical points and
critical points in the presence of disorder with specific focus on
the 2D Kitaev model which provides an explicit realization of both
the cases. First, we derive a generic model-independent expression
for the scaling of $n$ and $Q$ for such dynamics which takes a
$d$-dimensional system from a gapped phase to the vicinity of an
anisotropic critical point. We consider a scenario where the energy
gap $\Delta_{\vec k}$ vanishes as $k_i^z$ for $m$ momentum
components ($i=1..m$) and as $k_i^{z'}$ for the rest $d-m$
components ($i=m+1..d$) with $z' \ge z$ at the critical point and
show that the time-evolution of the Hamiltonian parameter
$\lambda(t)$, which brings the system at the critical point at
$t=0$, leads to novel scaling laws for $n$ and $Q$:
\begin{eqnarray}
n \sim \tau^{-[m + (d-m)z/z']\nu \alpha/(z\nu \alpha +1)},
\nonumber\\
Q \sim \tau^{-[(m+z)+ (d-m)z/z']\nu \alpha/(z\nu \alpha +1)}.
\end{eqnarray}
Our results reproduce their well-known counterparts for the
isotropic case ($z=z'$) as special cases. We also show, by exact
analytical solution for linear time evolution ($\alpha=1$), that the
two-dimensional (2D) Kitaev model, in the absence of disorder,
provides an explicit realization of the scaling laws mentioned above
with $d=z'=2$ and $m=\nu=z=1$ leading to $n \sim \tau^{-3/4}$ and
$Q\sim \tau^{-5/4}$. We also corroborate the scaling laws mentioned
above by numerical studies of the Kitaev model for arbitrary
power-law time evolution. Second, we compute all independent
multispin correlation function of the Kitaev model subsequent to a
slow linear ramp which takes the system from a gapped phase to the
vicinity of the anisotropic critical point, demonstrate their
anisotropic nature, and discuss their spatial dependence. Third, we
study non-equilibrium slow linear dynamics of the disordered Kitaev
model where disorder is introduced via random choice of the fields
$D_{\vec n}$ in the Fermionic representation of the model, and
show, by explicit numerical calculation, that the defect production
for such a dynamics obeys a different scaling law compared to its
disorder free counterpart: $n \sim \tau^{-1/2}$ and $Q\sim
\tau^{-1}$ ($Q\sim \tau^{-1/2}$) for a quench ending in the gapless
(gapped) phase. We provide a qualitative explanation for such defect
production.

The organization of the rest of the work is as follows. In Sec.\
\ref{anis1}, we discuss scaling laws for defect density and residual
energy for dynamics near an anisotropic critical point in the absence of
disorder and show that the 2D Kitaev model constitutes an example of
such a critical point. This is followed by Sec.\ \ref{corre1} where
we compute the equal-time correlation function of the 2D Kitaev
model following such a dynamics and discuss its spatial structure.
In Sec.\ \ref{dis1}, we discuss defect production in the disordered
Kitaev model. Finally we provide a discussion of our results and
conclude in Sec.\ \ref{conc1}.

\section {Anisotropic critical points}
\label{anis1}

\begin{figure}[tbp]
\begin{center}
\includegraphics[width=6cm,clip]{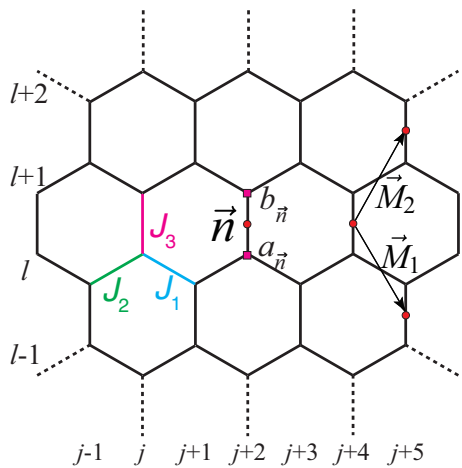}
\end{center}
\caption{Schematic representation of the Kitaev model on a honeycomb
lattice. The bonds $J_1$, $J_2$ and $J_3$ shows nearest neighbor
couplings between $x$, $y$ and $z$ components of the spins
respectively. $\vec{n}$ represents the position vector of the
midpoint of each vertical bond (unit cell). The vectors $\vec{M}_1$
and $\vec{M}_2$ are spanning vectors of the lattice. In the
Fermionic representation of the model, the Majorana Fermions
$a_{\vec{n}}$ and $b_{\vec{n}}$ sit at the bottom and top sites
respectively of the vertical bond with center coordinate $\vec{n}$
as shown. } \label{fig1}
\end{figure}
We begin with the study of slow dynamics in the Kitaev model
\cite{kitaev1,sengupta1}. The Hamiltonian for this model,
schematically represented in Fig.\ \ref{fig1}, is
given by
\begin{eqnarray}
H_K &=& \sum_{j+l={\rm even}} (J_1 \tau_{j,l}^x \tau_{j+1,l}^x + J_2
\tau_{j-1,l}^y \tau_{j,l}^y  + J_3 \tau_{j,l}^z \tau_{j,l+1}^z),
\label{hamkit1}
\nonumber\\
\end{eqnarray}
where $\vec \tau_{jl} = (\tau_{jl}^x,\tau_{jl}^y,\tau_{jl}^z)$
denote Pauli matrices at the site $(j,l)$ of the honeycomb lattice,
$J_1$, $J_2$, and $J_3$ represent nearest-neighbor couplings between
$x$, $y$ and $z$ components of the spins respectively. It is
well-known that $H_K$ can be represented in terms of Fermionic
fields by a straightforward Majorana transformation: $a_{jl} =
\left( \prod_{i=-\infty}^{j-1} ~\tau_{il}^z \right) ~ \tau_{jl}^y
~~{\rm for}~{\rm ~ even }~ j+l$ and $ b_{jl} = \left(
\prod_{i=-\infty}^{j-1} ~\tau_{il}^z \right) ~\tau_{jl}^x ~~ {\rm
for}~{\rm ~odd }~ j+l$ \cite{sengupta1}. This leads to the Fermionic
Hamiltonian
\begin{eqnarray}
H_{F} &=& i ~\sum_{\vec n} ~[J_1 ~b_{\vec n} a_{{\vec n} - {\vec
M}_1} ~+~ J_2 ~ b_{\vec n} a_{{\vec n} + {\vec M}_2} \nonumber \\
&& + J_3 D_{\vec n} ~b_{\vec n} a_{\vec n}],  \label{ham2}
\label{hamkit-ab}
\end{eqnarray}
where $\vec n = {\sqrt 3} {\hat i} ~n_1 + (\frac{\sqrt 3}{2} {\hat
i} + \frac{3}{2} {\hat j} ) ~n_2$ denote the midpoints of the
vertical bonds. Here $n_1, n_2$ run over all integers so that the
vectors $\vec n$ form a triangular lattice whose vertices lie at the
centers of the vertical bonds of the underlying honeycomb lattice.
The Majorana Fermions $a_{\vec n}$ and $b_{\vec n}$ sit at the bottom
and top sites respectively of the bond labeled $\vec n$. The
vectors ${\vec M}_1 = \frac{\sqrt 3}{2} {\hat i} - \frac{3}{2} {\hat
j}$ and ${\vec M}_2 = \frac{\sqrt 3}{2} {\hat i} + \frac{3}{2} {\hat
j}$ are spanning vectors for the lattice, and $D_{\vec n}$ can take
the values $\pm 1$ independently for each $\vec n$. The crucial
point that makes the solution of Kitaev model feasible is that
$D_{\vec n}$ commutes with $H_F$, so that all the eigenstates of
$H_{F}$ can be labeled by specific values of $D_{\vec n}$. It is
well-known that the ground state of the model corresponds to
$D_{\vec n}=1$ on all links \cite{kitaev1}.

For $D_{\vec n}=1$, Eq. (\ref{ham2}) can be diagonalized as
\begin{eqnarray}
H_F = \sum_{\vec k} \psi_{\vec k}^\dagger H_{\vec k} \psi_{\vec k},
\label{hamkit2}
\end{eqnarray}
where $\psi_{\vec k}^\dagger =(a_{\vec k}^\dagger, ~ b_{\vec
k}^\dagger)$ are Fourier transforms of $a_{\vec n}$ and $b_{\vec
n}$, the sum over $\vec k$ extends over half the Brillouin zone (BZ)
of the triangular lattice formed by the vectors $\vec n$, and
$H_{\vec k}$ can be expressed in terms of the Pauli matrices
$\sigma^i$ in particle-hole space as
\begin{eqnarray}
H_{\vec k} &=& 2 [J_1 \sin ({\vec k} \cdot {\vec M}_1)-J_2 \sin
({\vec k} \cdot {\vec M}_2)] \sigma^1 \nonumber\\
&& +2 [J_3 + J_1 \cos ({\vec k} \cdot {\vec M}_1) + J_2 \cos ({\vec
k} \cdot {\vec M}_2)] \sigma^2. \label{hamkit3}
\end{eqnarray}
The spectrum consists of two bands with energies $E_{\vec k}^\pm =
\pm E_{\vec k}$ \cite{sengupta1}, where
\begin{eqnarray}
E_{\vec k} &=& 2 [\{J_1 \sin ({\vec k} \cdot {\vec M}_1) - J_2 \sin
({\vec k} \cdot {\vec M}_2) \}^2  \nonumber\\
&& + \{ J_3 + J_1 \cos ({\vec k} \cdot {\vec M}_1) + J_2 \cos ({\vec
k} \cdot {\vec M}_2) \}^2 ]^{1/2}. \label{enkit1}
\end{eqnarray}
For $|J_1-J_2|\le J_3 \le J_1+J_2$, the bands touch each other, and
the energy gap $\Delta_{\vec k} = E_{\vec k}^+ - E_{\vec k}^-$
vanishes for special values of $\vec k$ leading to a gapless phase.
In particular we note that for $J_1=J_2=1$ and $J_3=2$, the gap
vanishes at $\vec k_c=(2\pi/\sqrt{3},0)$ and around this point
$\Delta_{\vec k} \sim k_y$ and $\Delta_{\vec k} \sim k_x^2$. Thus
this critical point constitutes an example of an anisotropic
critical point with $z=m=1$ and $d=z'=2$. We note that such an
anisotropic scaling occurs for any non-zero value of $J_1$ and $J_2$
at $J_3=(J_1+J_2)$.

We now consider a dynamics in this model $J_3(t) =(J_1+J_2 - J
t/\tau)$ from $t=-\infty$ to $t=0$ at a fixed rate $1/\tau$ which
brings the system from a gapped phase to the anisotropic critical
point at $\vec k_c$. Although this quench problem can be solved
for any $J_1$ and $J_2$, we shall fix $J_1=J_2=J$ for simplicity and
scale all energies (times) by $J$ ($\hbar/J$) in the subsequent
analysis. This choice does not change the scaling properties which
we seek. Also, to study the time evolution of the system, we note
that after an unitary transformation $U= \exp(-i \sigma^1 \pi/4)$,
we obtain $H_F = \sum_{\vec k} \psi_{\vec k}^{'\dagger} H'_{\vec k}
\psi'_{\vec k}$, where $H'_{\vec k} = U H_{\vec k} U^\dagger$ is
given by
\begin{eqnarray}
H'_{\vec k} = 2[ (g_{\vec k} -t/\tau) \sigma^3 + \alpha_{\vec k}
\sigma^{1} ], \label{hamkit5}
\end{eqnarray}
where $\alpha_{\vec k}= \sin ({\vec k} \cdot {\vec M}_1) - \sin
({\vec k} \cdot {\vec M}_2)$ and $g_{\vec k} = 2+ \cos ({\vec k}
\cdot {\vec M}_1) + \cos ({\vec k} \cdot {\vec M}_2)$. Hence the
off-diagonal elements of $H'_{\vec k}$ remain time independent, and
the quench problem reduces to a Landau-Zener problem for each $\vec
k$.

The state of the system after the quench at $t=0$ can be found by
solving the Landau-Zener problem at each $\vec k$ with the initial
condition $\psi_{\vec k}^G (t=-\infty)=|1\rangle= (0,1)^T$ for all
$\vec k$. After some algebra, one obtains for a given $\vec k$ and
at $t=0$ \cite{lzpapers1}
\begin{eqnarray}
|\psi_{\vec k}\rangle^d &=&  e^{-\pi \alpha_{\vec k}^2 \tau/4} \Big(
e^{3i\pi/4} D_{\mu_{\vec k}}(\nu_{\vec k}) |1\rangle
\nonumber\\
&& + \alpha_{\vec k} \sqrt{\tau} D_{\mu_{\vec k}-1}(\nu_{\vec k}
)|0\rangle \Big),\label{wavef1}
\end{eqnarray}
where $\nu_{\vec k}= 2 i g_{\vec k} \sqrt{\tau} \exp (- i \pi/4)$,
$\mu_{\vec k}= -i\alpha_{\vec k}^2 \tau$ and $D_{\mu}$ are parabolic
cylinder functions. The excited state at $t=0$, solved by
diagonalizing $H'_{\vec k}(t=0)$, yields, for a given $\vec k$, $
|\psi_{\vec k}^+ \rangle = ((E_{\vec k}^+ -2g_{\vec k})|1\rangle + 2
\alpha_{\vec k}|0\rangle)/{\mathcal D}_{\vec k}$, where ${\mathcal
D}_{\vec k}= [(E_{\vec k}^+ -2g_{\vec k})^2 + 4\alpha_{\vec
k}^2]^{1/2}$. Thus the probability of defect formation, given by
$p_{\vec k}=|\langle \psi_{\vec k}^+|\psi_{\vec k}\rangle^d|^2$, can
be obtained as
\begin{eqnarray}
p_{\vec k} &=& \frac{4 \alpha^2_{\vec k} e^{-\pi \alpha_{\vec k}^2
\tau/2}}{{\mathcal D}_{\vec k}^2} \Big|\alpha_{\vec k}\sqrt{\tau}
D_{\mu_{\vec k}-1}(\nu_{\vec k}) + \frac{E_{\vec k}^+ -2g_{\vec
k}}{2
\alpha_{\vec k}} \nonumber\\
&& \times  e^{-3i\pi/4} D_{\mu_{\vec k}}(\nu_{\vec k}) \Big |^2.
\label{prob1}
\end{eqnarray}
Since $\tau$ is large for slow dynamics, the contribution to the
defect formation comes from a small region near the critical point
where $\Delta_{\vec k}$ is sufficiently small for $\vec k \simeq
\vec k_c$. The density of defects can be thus estimated by expanding
$p_{\vec k}$ about $\vec k=\vec k_c$: $n \simeq \int d \delta k_x
\,d \delta k_y \, p_{\vec k= \vec k_c + \vec {\delta k}}$, where the
limits of integration can now be safely extended to infinity. To
compute this integral, we note that around $ \vec k= \vec k_c$, $
\alpha _{\vec {\delta k}} \simeq 3 \delta k_y$ and $ g_{\vec {\delta
k}} \sim 3(\delta k_x^2 + 3\delta k_y^2)/4$. Thus a redefinition of
variables $\delta k_x \to \delta k'_x = \delta k_x \tau^{1/4}$ and
$\delta k_y \to \delta k'_y =\delta k_y \tau^{1/2}$ allows us to
extract the $\tau$ dependence of the defect density
\begin{eqnarray}
n \simeq \int d \delta k_x \,d \delta k_y \, p_{\vec {\delta k}}, \,
\sim \tau^{-3/4} \int d \delta k'_x \,d \delta k'_y \, p_{\vec
{\delta k'}}.\label{defectscaling}
\end{eqnarray}
A similar analysis can be carried out for computation of residual
energy $Q = (2\pi)^{-2}\int d^2k p_{\vec k} \Delta_{\vec k}$. Here we
note that near the critical point $\vec k=\vec k_c$, $\Delta_{\vec k}
\simeq 4\sqrt{9 \delta k_y^2 + 9(\delta k_x^2 + 3 \delta
k_y^2)^2/16}$ and thus scale as $\tau^{-1/2}$. Thus one obtains
\begin{eqnarray}
Q \simeq \int d \delta k_x \,d \delta k_y \,\Delta_{\vec{\delta k}}
p_{\vec {\delta k}}, \, \sim \tau^{-5/4} \int d \delta k'_x \,d
\delta k'_y \,\Delta_{\vec{\delta k'}} p_{\vec {\delta k'}}.
\label{energyscaling} \nonumber\\
\end{eqnarray}
Eqs. (\ref{defectscaling}) and (\ref{energyscaling}) show that $n
\sim \tau^{-3/4}$ and $Q \sim \tau^{-5/4}$ at the critical point.
These scaling laws do not conform to the predictions of earlier
works on defect production during passage through isotropic quantum
critical points \cite{pol2} or critical surfaces \cite{sengupta1};
their origin lies in the anisotropic scaling of $\delta k_x$ and
$\delta k_y$ with the quench time $\tau$.

To generalize these results for arbitrary $d$-dimensional anisotropic
critical points, where the energy gap $\Delta_{\vec k} \sim k_i^z$
for $m$ directions and $\sim k_i^{z'}$ for $d-m$ directions, we
provide a simple phase space argument as first proposed in Ref.\
\onlinecite{zurek1}. We consider a general power-law quench with
$\lambda(t) = \lambda_0 |t/\tau|^{\alpha} {\rm Sgn}(t)$ which starts
at $t=-\infty$ and reaches the critical point at $t=0$. We first
note that the adiabaticity condition breaks down when the rate of
change of the energy gap become equivalent to the square of the gap:
$d \Delta_{\vec k}/dt \ge \Delta_{\vec k}^2$. Since $\Delta_{\vec k}
\sim \lambda^{z\nu} |t/\tau|^{z\nu \alpha}$, we find that the time
spent by the system in the non-adiabatic regime is given by $\hat{t}
\sim \tau^{z\nu \alpha/(z \nu \alpha +1)}$. The scaling of the
energy gap in this regime can thus be written as $\Delta_{\vec k}
\sim \tau^{-z\nu \alpha/(z\nu \alpha +1)}$. The phase space for
defect production is given by $\Omega_n \sim k_1 .. k_d$. Since
$\Delta_{\vec k} \sim k_i^z$ for $i=1..m$ and $k_i^{z'}$ for
$i=m+1..d$, we finally obtain
\begin{eqnarray}
n \sim \tau^{-(m + (d-m)z/z') \nu \alpha/(z \nu \alpha +1)}.
\label{nfinal}
\end{eqnarray}
A similar argument can also be presented for the residual energy. We
note that for $z\le z'$, the leading behavior of the energy gap near
the quantum critical point, where the defects are produced, is
$\Delta_{\vec k} \sim k_i^z$ for $1 \le i \le m$. Thus the phase
space for the residual energy production is $\Omega_Q \sim
\Delta_{\vec k} k_1.. k_d$ leading to a scaling of $Q$ as
\begin{eqnarray}
Q \sim \tau^{-[(m+z)+(d-m)z/z'] \nu \alpha/(z\nu \alpha +1)}.
\label{qfinal}
\end{eqnarray}
We note that the scaling laws, Eqs.\ (\ref{nfinal}) and
(\ref{qfinal}), reproduce their isotropic counterparts for $z=z'$
leading to $n \sim \tau^{-d\nu\alpha/(z\nu\alpha+1)}$ and $Q \sim
\tau^{-(d+z)\nu \alpha/(z\nu \alpha +1)}$ \cite{sen1,pol2,pol3}.
Also, the scaling of the Kitaev model for linear time evolution
elaborated in this work is reproduced for $d=z'=2$, and
$z=\nu=\alpha=1$ leading $n\sim \tau^{-3/4}$ and $Q \sim
\tau^{-5/4}$. Moreover, we note that the scaling of defect density
for a linear quench through a gapless surface can also be obtained
from Eq.\ (\ref{nfinal}) by noting that for such quenches the energy
gap depends only on the $m$ momenta components orthogonal to the
$d-m$ dimensional gapless surface. This can be represented by
putting $z' \to \infty$ (since $k_{\parallel} \sim \Delta_{\vec
k}^{1/z'}$) leading to the scaling law $n \sim \tau^{-m \nu/(z\nu
+1)}$ \cite{sengupta1}. Thus Eqs.\ (\ref{nfinal}) and (\ref{qfinal})
reproduce all earlier results on defect production for slow dynamics
across quantum critical lines and surfaces as special cases.
Finally, we would like to point out that the maximum values of these
exponents is $2$ which can be obtained by similar considerations as
in the cases of isotropic critical points \cite{pol3}.

\begin{figure}[tbp]
\begin{center}
\includegraphics[width=8cm,clip]{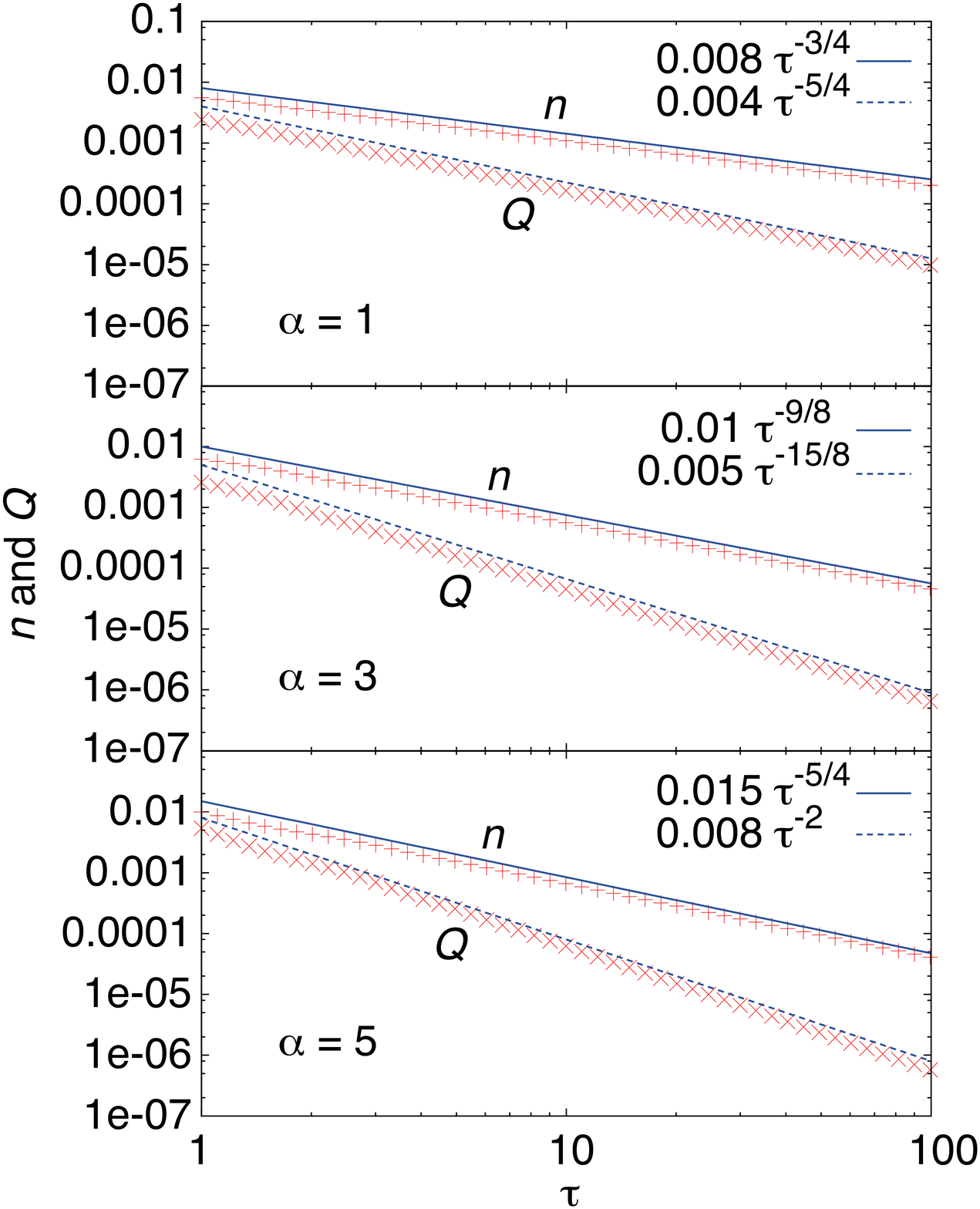}
\end{center}
\caption{Numerical results on the defect density $n$ and residual
energy $Q$. The time-dependent Schr\"{o}dinger equation is solved in
the momentum space for systems with size up to  $512\times 512$ unit
cells. The parameter $\alpha$ specifying the evolution of $J_3 =
2-|t/\tau|^{\alpha} {\rm Sgn}(t)$ is chosen as $\alpha=1$, $3$ and
$5$. The lines indicate the power laws expected from Eqs.\
(\ref{nfinal}) and (\ref{qfinal}) for $d=z'=2$ and $z=\nu=m=1$,
$n\sim\tau^{-3\alpha/[2(\alpha+1)]}$ and
$Q\sim\tau^{-5\alpha/[2(\alpha+1)]}$. The agreement between
curves obtained numerically and the corresponding power laws
is remarkable.
In all plots, $t$ varies from an initial value $t_{\rm in}=
-3\tau$ to a final value $t_{\rm f}=0$. } \label{fig2}
\end{figure}

To verify these scaling laws, we now study non-linear power-law
dynamics in the Kitaev model numerically. To this end, we again
restrict ourselves to $J_1=J_2=1$ and evolve $J_3(t)= (2 -
|t/\tau|^{\alpha} {\rm Sgn}(t))$ for $-\infty \le t \le 0$ so that
the anisotropic critical point is reached at $t=0$. The
corresponding time-dependent Hamiltonian is given by $H(\vec k; t) =
\sum_{\vec k} \psi_{\vec k}^{\dagger} \left[ (g_{\vec k} -
|t/\tau|^{\alpha} {\rm Sgn}(t)) \sigma^3 + \alpha_{\vec k} \sigma^1
\right] \psi_{\vec k}$. We solve the time-dependent Schr\"odinger
equation $i
\partial_t \psi_{\vec k} = H(\vec k;t) \psi(\vec k)$ numerically for
each $k$, compute $p_{\vec k}$, and use it to obtain the defect
density $ n = \int d^2 k p_{\vec k}$ and $Q = \int d^2k \Delta_{\vec
k} p_{\vec k}$ numerically as a function of $\tau$ and $\alpha$. The
plots of $n$ and $Q$ vs $\tau$ are shown in Fig.\ \ref{fig2} for
several representative values of $\alpha$. The lines in the figure
indicate the power laws expected from Eqs.\ \ref{nfinal} and
\ref{qfinal} ($n \sim \tau^{-3\alpha/[2(\alpha+1)]}$ and $Q\sim
\tau^{-5\alpha/[2(\alpha+1)]}$) for $d=z'=2$ and $z=\nu=m=1$. The
agreement between the numerical and theoretical results corroborates
the scaling theory proposed in this work.

\section{Correlation function}
\label{corre1}

In this section, we compute the independent correlation function for
the Kitaev model for linear time evolution. Since the model can be
represented by free Fermions, it is easy to see that the only
non-zero independent correlators are those between free Fermions
which are given by
\begin{eqnarray}
\langle O_{\vec r} \rangle &=& i \langle b_{\vec n} a_{\vec n +\vec
r} \rangle
\nonumber\\
&=& \frac{4i}{N_s} \sum_{\vec k} [\langle b_{\vec k}^{\dagger} a_{\vec
 k}\rangle \exp
(i\vec k \cdot \vec r) - {\rm h.c.} ],
\end{eqnarray}
where $\langle .. \rangle$ denotes expectation value with respect to
a direct product of states involving $\vec{k}$ only, ${\rm
h.c.}$ denotes hermitian conjugate, and $N_s$ is the number of
sites. After an unitary transformation $U=\exp(-i\sigma^1\pi/4)$,
we find
\begin{equation}
\langle O_{\vec{r}} \rangle = -\frac{2}{N}\sum_{\vec{k}} \langle
\psi_{\vec{k}}^{' \dagger}
[-\cos (\vec{k}\cdot\vec{r}) \sigma^3+\sin (\vec{k}\cdot\vec{r}) \sigma^1]
 \psi'_{\vec{k}}\rangle. \label{dcorr}
\end{equation}
The interpretation of these correlation functions in terms of the
original spin degrees of freedom have already been pointed out in
Ref.\ \onlinecite{sengupta1}. For $\vec r=0$, $\langle O_{\vec
r}\rangle $ represents correlations between $z$ components between
spins at the end of the vertical bond whose midpoint is denoted by
$\vec n$. For ${\vec r}\ne 0$, it represents correlation between
product of multiple spin operators which begins with $\tau^x$ or
$\tau^y$ on a $b$ or $a$ site at $\vec n=(j,l)$ and ends with
$\tau^x$ or $\tau^y$ on an $a$ or $b$ site at $\vec n +\vec
r=(j',l')$ with a string of $\tau^z$ operators living on sites in
between. Note that the Fermionic representation in terms of free
Fermions with $D_{\vec n}=1$ ensures that these multispin
correlation functions are the only non-zero independent spin
correlation functions of the model.
\begin{figure}[tbp]
\begin{center}
\includegraphics[width=7cm,clip]{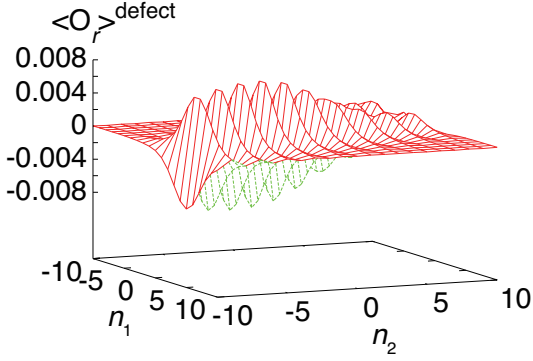}
\end{center}
\caption{Plot of the correlation function
$\langle O_{\vec r}\rangle^{defect}$ as a function
of $n_1$ and $n_2$ for linear quench of $J_3/J_1$ from $5$ to $2$
with $J_1=J_2=1$ and $\tau=5$.} \label{figcorr1}
\end{figure}

The ground state with a fixed $\vec{k}$ for $J_3 = 2$ is given by
$|\psi_{\vec{k}}^{-}\rangle =
((E_{\vec{k}}^{-}-2g_{\vec{k}})|1\rangle
+2\alpha_{\vec{k}}|0\rangle)/{\mathcal D}_{\vec{k}}^-$, where
${\mathcal D}_{\vec{k}}^- = [(E_{\vec{k}}^- -
2g_{\vec{k}})^2+4\alpha_{\vec{k}}^2]^{1/2}$. Noting that $|0\rangle$
and $|1\rangle$ are basis of $\psi'_{\vec{k}}$, one finds, from Eq.
(\ref{dcorr}), the correlation function of the ground state as
\begin{eqnarray}
 \langle O_{\vec{r}}\rangle^{G}
 &=& - \delta_{\vec{r 0}} + \frac{1}{A}\int d^2k
  \left[2\frac{4\alpha_{\vec{k}}^2}{({\mathcal D}_{\vec{k}}^-)^2}
   \cos(\vec{k}\cdot\vec{r}) \right. \nonumber \\
 && - \left.
   2\frac{2\alpha_{\vec{k}}(E_{\vec{k}}^- - 2g_{\vec{k}})}
   {({\mathcal D}_{\vec{k}}^-)^2}\sin(\vec{k}\cdot\vec{r}) \label{dcorr0}
\right],
\end{eqnarray}
where $A=4\pi/3\sqrt{3}$ is the area of half of the Brillouin zone.
As for the state after quench,
a straightforward calculation using Eq.\ (\ref{wavef1}) shows
\begin{eqnarray}
\langle O_{\vec r}\rangle^{d} &=& -\delta_{\vec r 0} + \frac{1}{A} \int
 d^2k e^{-\pi
\alpha_{\vec k}^2 \tau/2} \alpha_{\vec k} \sqrt{\tau} \Big[ 2
\alpha_{\vec k} \sqrt{\tau} \nonumber\\
&& \times |D_{\mu_{\vec k}-1}(\nu_{\vec k}) |^2 \cos(\vec k \cdot
\vec r) - \Big\{ e^{i 3\pi/4} D^{\ast} _{\mu_{\vec k}-1}(\nu_{\vec
k})  \nonumber\\
&& \times D_{\mu_{\vec k}} (\nu_{\vec k}) +{\rm c.c.} \Big\}
\sin(\vec k \cdot \vec r)\Big]. \label{dcorr1}
\end{eqnarray}
Note that $\langle O_{\vec{r}}\rangle^d$ reduces to $\langle
O_{\vec{r}}\rangle^G$ with $\tau\to\infty$. The correlation between
defects induced by non-adiabatic quench dynamics can be captured by
the deviation of $\langle O_{\vec{r}}\rangle^d$ from $\langle
O_{\vec r}\rangle^G$. Thus we define the defect correlation function
by
\begin{equation}
\langle O_{\vec{r}}\rangle^{\rm defect} = \langle O_{\vec
r}\rangle^d - \langle O_{\vec r}\rangle^G . \label{dcorr2}
\end{equation}
The nature of the spatial dependence of the defect correlation
function for slow dynamics (large $\tau$) can be qualitatively
understood for $J_1=J_2$ from Eq.\ (\ref{dcorr1}). To this end, we
first separate the contribution to $\langle O_{\vec r}\rangle^{d}$
which comes from around $\alpha_{\vec k} \simeq 1/\tau$ from those
coming from other regions in the $\vec k$-space. For estimating the
latter contribution, we consider $\tau \gg 1$ so that for
$\alpha_{\vec k}, g_{\vec k} \ne 0$, $|\mu_{\vec k}|, |\nu_{\vec k}|
\to \infty$. We then note that the following identities for $D$
holds in the limit $b \to \infty$ with arbitrary ratio $a/b$
\begin{eqnarray}
e^{-\pi b^2/4} D_{-i b^2 -1}(ae^{i\pi/4})\simeq \sin(\theta)e^{-i
(\eta +\pi/4)}/b, \nonumber\\
e^{-\pi b^2/4} D_{-i b^2}(ae^{i\pi/4})\simeq \cos(\theta)e^{-i \eta
}, \label{id1}
\end{eqnarray}
where $\theta$ and $\eta$ are defined through the relations
\begin{eqnarray}
\cos(\theta)(\sin (\theta)) &=& \sqrt{ [1 +(-) a/(2
\sqrt{b^2+a^2/4})]/2}, \nonumber\\
\eta &=& -b^2/2 + b^2 \ln(a/2+\sqrt{b^2+a^2/4})\nonumber\\
&& +a \sqrt{b^2+a^2/4}/2.  \label{id2}
\end{eqnarray}
\begin{figure}[tbp]
\begin{center}
\includegraphics[width=7cm,clip]{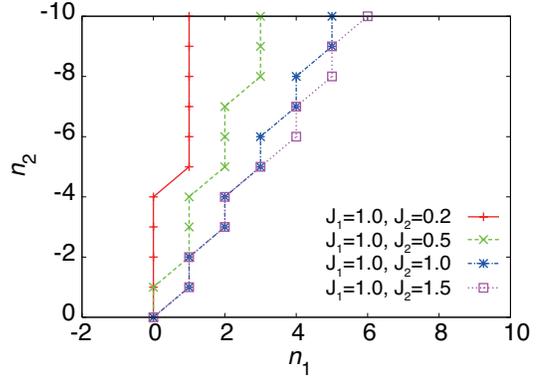}
\end{center}
\caption{Plot of the peak positions of $\langle O_{\vec
r}\rangle^{defect}$ in the $n_1-n_2$ plane for $J_1=1$ and several
representative values of $J_2$. For each of these cases, the quench
starts at $J_3/J_1=5$ and ends at the anisotropic critical point.
Note that the axis of $n_2$ is upside down.} \label{figcorr2}
\end{figure}
Identifying $b= \alpha_{\vec k} \sqrt{\tau}$ and $a = 2 g_{\vec k} \tau$
and substituting Eqs.\ (\ref{id1}) and  (\ref{id2}) in Eq.\
(\ref{dcorr1}), we find, after some straightforward algebra, that the
integrand of Eq.\ (\ref{dcorr1}) reduces to that of Eq.\ (\ref{dcorr0})
for all $\vec k$ except those for which $\alpha_{\vec k} \sqrt{\tau} \simeq
1$. Thus, one finds that in this limit, the main contribution to
$\langle O_{\vec{r}}\rangle^{\rm defect}$ comes from around the line
$\alpha_{\vec k} \simeq 1/\sqrt{\tau}$. For large $\tau$, this is
infinitesimally close to the line $\sin(\vec k \cdot \vec M_1)=
\sin( \vec k \cdot \vec M_2)$. In this region of $k$ space,
$|D_{\mu_{\vec k}-1}(\nu_{\vec k}) |^2 \simeq |D_{i-1}(\nu_{\vec
k})|^2$ which, for large $\tau$ and $J_3=2$, is a sharply peaked
function for $g_{\vec k} \simeq 0$ which occurs at $\vec k = \vec
k_c$. Also, for $\vec k \simeq \vec k_c$, it can be easily checked
that $[D^{\ast} _{\mu_{\vec k}-1}(\nu_{\vec k}) D_{\mu_{\vec k}}
(\nu_{\vec k}) +{\rm h.c.}] \ll |D_{\mu_{\vec k}-1}(\nu_{\vec k})
|^2$, so that the major contribution to $\langle O_{\vec
r}\rangle^{\rm defect}$ comes from the coefficient of the $\cos(\vec
k \cdot \vec r)$ in the integrand. Using these observations and
expressing $\vec r = (\sqrt{3}(n_1+n_2/2), 3n_2/2)$, one can
estimate the spatial dependence of the correlation function
in the same line as in Ref.\
\onlinecite{sengupta1}. In particular, the maxima of the correlation
function is expected to occur along the maxima of $\cos(\vec k_c
\cdot \vec r)$ {\it i.e.} along the line $n_1 +n_2/2=0$ in the
$n_1-n_2$ plane. Away from this line, as shown in Ref.\
\onlinecite{sengupta1}, $\langle O_{\vec{r}}\rangle^{\rm defect}$ is
expected to decay exponentially as a function of $r$ with a
characteristic decay length $\sim \sqrt{\tau}$.

A plot of $\langle O_{\vec r}\rangle^{\rm defect}$ as a function of
$n_1$ and $n_2$, obtained by numerical evaluation of Eq.\
(\ref{dcorr2}) are shown in Fig.\ \ref{figcorr1} for $J_1=J_2=1$,
corroborates the above-mentioned discussion. We find that $\langle
O_{\vec r}\rangle^{\rm defect}$ peaks along the $n_1=-n_2/2$ line
and decays to zero as we move away from this line. The decay length
in the $n_1-n_2$ depends on $\tau$; for larger $\tau$ we have a
sharper decay. The slope of the line along which $\langle O_{\vec
r}\rangle^{\rm defect}$ peaks in the $n_1-n_2$ plane changes with
$J_1/J_2$ since ${\vec k}_c$ depends on this ratio. This can be seen
from Fig.\ \ref{figcorr2} which plots the position of the peaks of
the correlation functions for several representative values of
$J_1/J_2$. The analysis of the preceding paragraph can be easily
extended to these cases in the same line as in Ref.
\onlinecite{sengupta1} and is found to match the numerical results
for all $J_1/J_2$.

\section {Disordered Kitaev model}
\label{dis1}

In this section, we study the dynamics of Kitaev model given by Eq.\
(\ref{hamkit2}) with a random configuration of $D_{\vec{n}}$, namely
for random assignment of values $\pm 1$ to the link variables
$D_{\vec{n}}$. The dynamics is incorporated in the form of a
power-law evolution of $J_3$  as in Sec.\ \ref{anis1}.

The Hamiltonian (Eq.\ (\ref{hamkit-ab})) can be expressed
using the real-space Fermion operators
$\alpha_{\vec{n}}=\frac{1}{2}(b_{\vec{n}} - ia_{\vec{n}})$
at position $\vec{n}$ as
\begin{eqnarray}
 H_F &=& \sum_{\vec{n}} J_1\left(\alpha_{\vec{n}} +
            \alpha_{\vec{n}}^{\dagger}\right)
 \left(\alpha_{\vec{n}-\vec{M}_1} -
  \alpha_{\vec{n}-\vec{M}_1}^{\dagger}\right) \nonumber\\
 && +
 J_2\left(\alpha_{\vec{n}} + \alpha_{\vec{n}}^{\dagger}\right)
 \left(\alpha_{\vec{n}+\vec{M}_2} -
  \alpha_{\vec{n}+\vec{M}_2}^{\dagger}\right) \nonumber\\
&& + J_3 D_{\vec{n}} (1 - 2
\alpha_{\vec{n}}^{\dagger}\alpha_{\vec{n}}). \label{hamdis1}
\end{eqnarray}
The first two terms represent hopping and pair-creation and
annihilation of the Fermions while the third term induces a random
local potential. For $J_3\gg J_{1,2}$, the third term dominates and
the ground state of the system is composed of localized states of
the Fermion. In contrast for $J_3=0$, the ground state is clearly
delocalized. We now show numerically that a quantum phase transition
takes place in between these two limits at $J_3=J_{3,c}$. Note that
the existence of a sharp transition in the presence of the disorder
is consistent with the Harris criteria $\nu d \ge 2$ since for the
Kitaev model $d=2$ and $\nu=1$.

The Hamiltonian, Eq.\ (\ref{hamdis1}), is written in a quadratic
form as $H =
\psi^{\dagger}M\psi$ with  $\psi^{\dagger} =
(\alpha_{\vec{n}_1}^{\dagger},
 \alpha_{\vec{n}_2}^{\dagger}, \cdots, \alpha_{\vec{n}_N}^{\dagger},
 \alpha_{\vec{n}_1},  \alpha_{\vec{n}_2},
 \cdots, \alpha_{\vec{n}_N})$, where
$N$ is the number of vertical bonds (unit cells) in the system, and $M$
is a $2N\times 2N$ matrix given by
\begin{equation}
 M = \frac{1}{2}\left[
 \begin{array}{@{\,}cc@{\,}}
 A & B \\
 B^{\rm T} & -A \end{array}
\right],
\end{equation}
with
\begin{eqnarray}
 A_{\vec{n}_i-\vec{M}_1, \vec{n}_i} &=& A_{\vec{n}_i,\vec{n}_i-\vec{M}_1}
  = - J_1 , \nonumber\\
  A_{\vec{n}_i+\vec{M}_2, \vec{n}_i} &=& A_{\vec{n}_i,\vec{n}_i+\vec{M}_2}
  = - J_2 , \nonumber\\
  A_{\vec{n}_i, \vec{n}_i} &=&  2 J_3 D_{\vec{n}_i}, \nonumber\\
 B_{\vec{n}_i-\vec{M}_1, \vec{n}_i} &=& - B_{\vec{n}_i,\vec{n}_i-\vec{M}_1}
  = - J_1 , \nonumber\\
 B_{\vec{n}_i+\vec{M}_2, \vec{n}_i} = - J_2  &=& - B_{\vec{n}_i,\vec{n}_i+\vec{M}_2}.
\end{eqnarray}
All other elements of $A$ and $B$ are zero.
The matrix $M$ is diagonalized by a unitary matrix,
\begin{equation}
 U = \left[
 \begin{array}{@{\,}cc@{\,}}
 u & v^{\ast} \\
 v & u^{\ast} \end{array}
\right] ,
\end{equation}
as $U^{\dagger} M U = D$,
where $D$ is a diagonal matrix.
We note that the form
of $M$ necessitates that if $\epsilon_{\mu}$ is an eigenvalue of
$M$, so is $-\epsilon_{\mu}$. We hereafter suppose $\epsilon_{\mu} > 0$
and choose $U$ so that
$\epsilon_{\mu}$ ($\mu = 1,2,\cdots,N$) enter upper half diagonal
elements of $D$.
Defining a fermion operator as
\begin{equation}
 \gamma_{\mu} = \sum_{\vec{n}}
  u^{\ast}_{\vec{n},\mu}\alpha_{\vec{n}}
  + v^{\ast}_{\vec{n},\mu}\alpha^{\dagger}_{\vec{n}} ,
\end{equation}
the diagonalized Hamiltonian is written as
\begin{equation}
 H_F = \sum_{\mu=1}^N\epsilon_{\mu}(2\gamma^{\dagger}_{\mu}\gamma_{\mu} - 1).
\label{HFdiag}
\end{equation}
The ground-state energy is given by $E_g = -
\sum_{\mu=1}^N\epsilon_{\mu}$.
The energy gap from the ground state to the first excited
state is thus given by $\Delta = 2\epsilon_1$ where $\epsilon_1$ is
the smallest positive eigenvalue.

With this observation, we now compute the gap $\Delta$ numerically
for finite sizes and obtain the distribution of gaps by changing the
configuration of $\{D_{\vec{n}}\}$.
\begin{figure}[tbp]
\begin{center}
\includegraphics[width=7cm,clip]{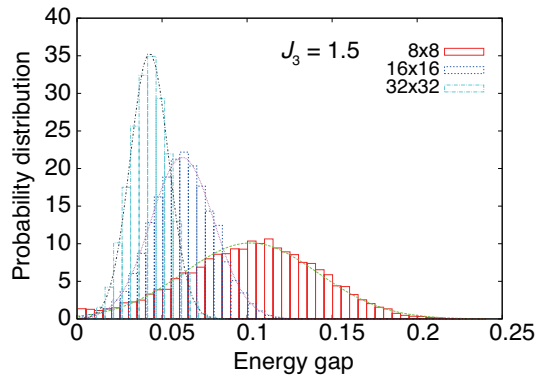}
\end{center}
\caption{Probability distribution of excitation gaps at $J_3=1.5$.
$10000$ instances of $\{D_{\vec{n}}\}$ are generated and for each of
them we obtained the excitation gap $2\epsilon_1$ by numerically
diagonalizing the matrix $M$. } \label{fig:DistGap1.5}
\end{figure}
We find that the property of the distribution of gaps is
qualitatively different for $J_3\lsim 1.5$ and $J_3\gsim 1.5$. Let
us first consider the case with $J_3 = 1.5$ for which the
distribution of the gaps is shown in Fig.\ \ref{fig:DistGap1.5} for
several system sizes. The shown distribution allows a Gaussian fit:
$N(\Delta)=(2\pi\sigma^2)^{-1/2}e^{-(\Delta-\bar{\Delta})^2/2\sigma^2}$
using the average $\overline{\Delta}$ and the variance $\sigma^2 =
\overline{\Delta^2} - \overline{\Delta}^2$, where $\bar{}$ stands
for the average over the random configuration of $\{D_{\vec{n}}\}$.
\begin{figure}[tbp]
\begin{center}
\includegraphics[width=7cm,clip]{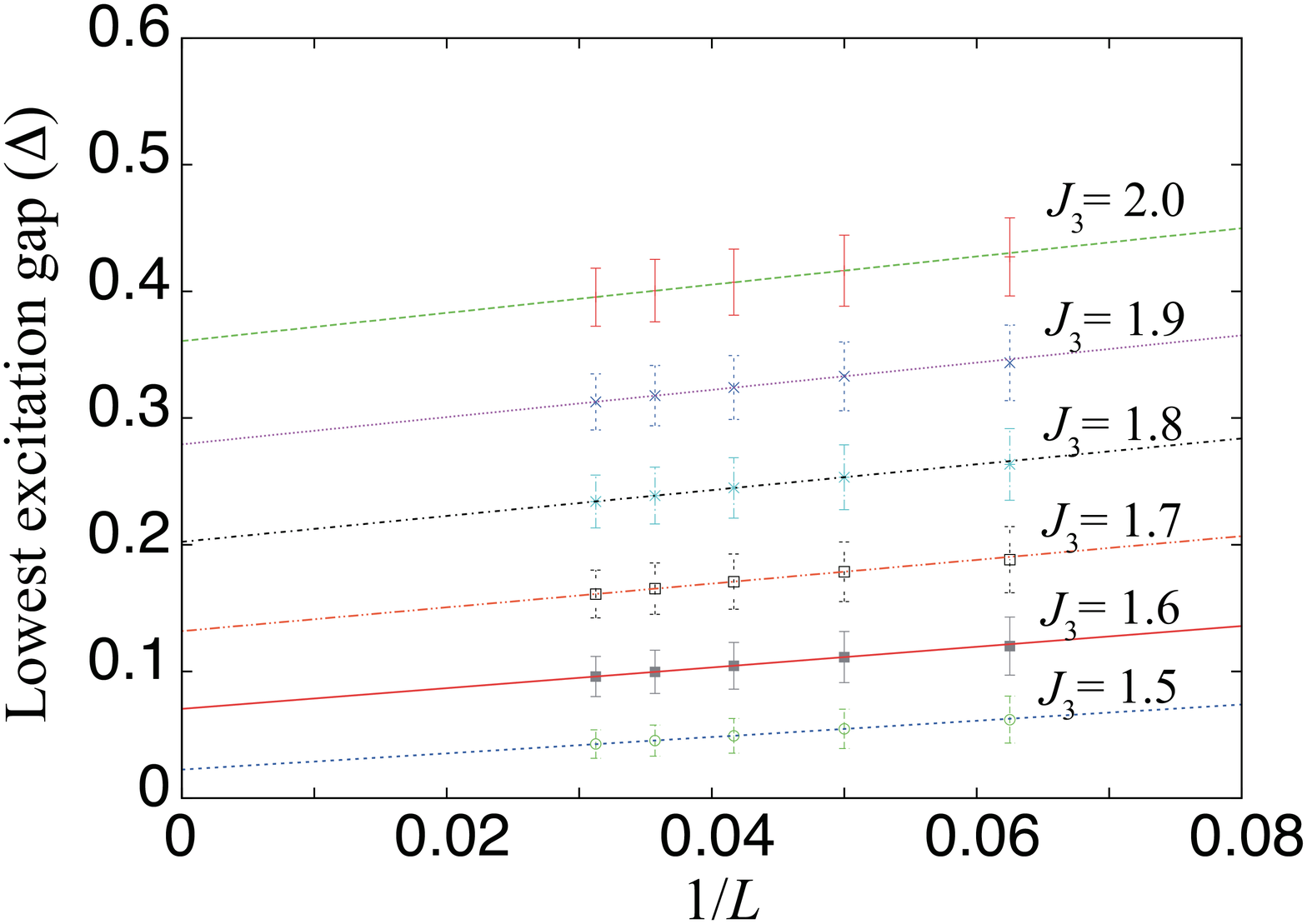}
\end{center}
\caption{Finite size scaling of the average of gaps for $J_3 = 1.5$,
$1.6$, $1.7$, $1.8$, $1.9$, and $2.0$. We find that the average of
gaps is scaled by $1/L$, where $L$ is the length of the system
($N=L^2$). } \label{fig:SizeScaling1}
\end{figure}
Figure \ref{fig:SizeScaling1} shows the size scaling of $\bar
\Delta$ for several $J_3\geq 1.5$. We find ${\bar \Delta}$ scales
linearly with $1/L$. Since the variance of gaps tends to vanish for
$L\to\infty$, one can estimate the gap in the thermodynamic limit
$\Delta_{\infty}$ by extrapolating the fitting line of
$\bar{\Delta}$ for $1/L\to 0$. Such a behavior of $\Delta_{\infty}$
is to be contrasted with that for $J_3=1$ as shown in Fig.
\ref{fig:ScaleGapProb}. For $J_3=1$, we find that the probability
distribution of gaps scales as $\Delta \sim 1/L^2$ as seen from the
collapse of the data for several system sizes (Fig.\
\ref{fig:ScaleGapProb}). The difference in behavior of ${\bar
\Delta}$ can be further understood by plotting $\Delta_{\infty}$ for
several values of $J_3$. This is shown in Fig. \ref{fig:Gap-J3}. We
find that for $J_3\lsim 1.4$ the gap vanishes, while it increases
linearly with $J_3$ for $J_3\ge 1.5$. The position of the critical
point $J_{3,c}$ can therefore be estimated to be around $1.5$.
Moreover, the gap $\Delta$ increases as $\Delta\propto |J_3 -
J_{3,c}|$ for $J_3\ge 1.5$ leading to $z\nu = 1$ for the transition.

Having established the presence of a quantum critical point in the
disordered Kitaev model, we now study the dynamical behavior during
slow non-adiabatic linear time evolution $J_3(t)=-Jt/\tau$ which
takes the system from a gapped region ($J_{3}=5$) either to a
gapless region ($J_3=0$) or to a gapped region passing through the
gapless region ($J_3=-5$). In order to obtain quantities of
interest, we switch to the Heisenberg picture \cite{barouch,caneva}
and introduce the time-evolution operator $U(t)$, $|\Psi(t)\rangle =
U(t)|\Psi(t_{\rm in })\rangle$, where $t_{\rm in}$ denotes the
initial time. The operator $\alpha_{\vec{n}}$ in the Heisenberg
picture is denoted by $\alpha^{\rm H}_{\vec{n}}(t) =
U^{\dagger}(t)\alpha_{\vec{n}}U(t)$. Computing the commutator of
$\alpha_{\vec{n}}$ and $H_F$ expressed by Eq.\ (\ref{hamdis1}), the
Heisenberg equation of motion for $\alpha^{\rm H}_{\vec{m}}$ is
given by
\begin{equation}
 i \frac{d}{dt} \alpha^{\rm H}_{\vec{m}}(t)
  = \sum_{\vec{n}}\left(A_{\vec{m},\vec{n}}\alpha_{\vec{n}}^{\rm H}(t)
 + B_{\vec{m},\vec{n}}\alpha^{{\rm H} \dagger}_{\vec{n}}(t)\right).
\label{eqalpha}
\end{equation}
\begin{figure}[tbp]
\begin{center}
\includegraphics[width=7cm,clip]{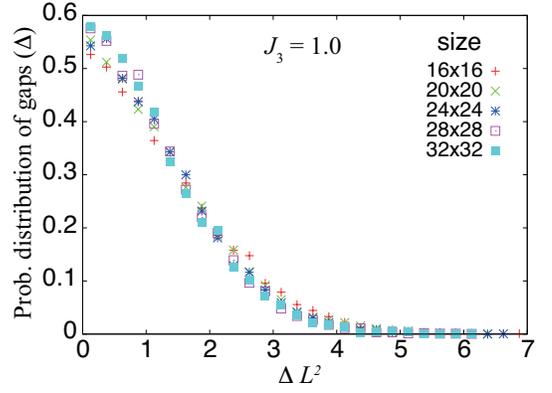}
\end{center}
\caption{The probability distribution of gaps at $J_3 = 1.0$.
Horizontal axis is the excitation gap multiplied by $L^2$. The
curves with different size almost collapse, meaning that the
distribution of gaps is given by a function of $\Delta L^2$ and the
gap vanishes as $1/L^2$ with increasing $L$. }
\label{fig:ScaleGapProb}
\end{figure}
We define matrices $u_{\vec{m},\nu}(t)$ and $v_{\vec{m},\nu}(t)$
by an expansion of $\alpha^{\rm H}_{\vec{m}}(t)$ by
$\gamma_{\nu,{\rm in}}$, operators which diagonalize the Hamiltonian
at initial time $t_{\rm in}$ (see Eq.\ (\ref{HFdiag})):
\begin{equation}
 \alpha^{\rm H}_{\vec{m}}(t) = \sum_{\nu}
  \left(u_{\vec{m},\nu}(t)\gamma_{\nu,{\rm in}}
  + v^{\ast}_{\vec{m},\nu}(t)\gamma^{\dagger}_{\nu,{\rm in}}\right).
\label{BdG1}
\end{equation}
Substituting this expansion for $\alpha^{\rm H}$'s in Eq.\ (\ref{eqalpha}),
one obtains equations of motion for $u_{\vec{m},\nu}(t)$ and
$v^{\ast}_{\vec{m},\nu}(t)$:
\begin{eqnarray}
 i \frac{d}{dt} u_{\vec{m},\nu}(t) &=&
  \sum_{\vec{n}}A_{\vec{m},\vec{n}} u_{\vec{n},\nu}(t)
  + B_{\vec{m},\vec{n}} v_{\vec{n},\nu}(t), \\
  i \frac{d}{dt} v^{\ast}_{\vec{m},\nu}(t) &=&
  \sum_{\vec{n}}A_{\vec{m},\vec{n}} v^{\ast}_{\vec{n},\nu}(t)
  + B_{\vec{m},\vec{n}} u^{\ast}_{\vec{n},\nu}(t).
\end{eqnarray}
The initial conditions for $u_{\vec{m},\nu}(t)$ and
$v_{\vec{m},\nu}(t)$ are written as $u_{\vec{m},\nu}(t_{\rm
in})=u_{\vec{m},\nu,{\rm in}}$ and $v_{\vec{m},\nu}(t_{\rm
in})=v_{\vec{m},\nu,{\rm in}}$, where $u_{\rm in}$ and $v_{\rm in}$
are block matrices of $U$ diagonalizing $M$ at initial time. To
obtain the expressions of $n$ and $Q$ at final time $t_{\rm f}$, we
introduce notations $u_{\rm f}$, $v_{\rm f}$, $\epsilon_{\mu,{\rm
f}}$, and $\gamma_{\mu,{\rm f}}$ so that $H_F(t_{\rm f}) =
\sum_{\mu}\epsilon_{\mu,{\rm f}} (2\gamma^{\dagger}_{\mu,{\rm
f}}\gamma_{\mu,{\rm f}} - 1)$, where
\begin{equation}
 \gamma_{\mu,{\rm f}} = \sum_{\vec{n}}
  u^{\ast}_{\vec{n},\mu,{\rm f}}\alpha_{\vec{n}}
  + v^{\ast}_{\vec{n},\mu,{\rm f}}\alpha^{\dagger}_{\vec{n}}.
\label{gammaf}
\end{equation}
The density of excitation $n$ and the residual energy $Q$ can now
be defined by
\begin{eqnarray}
 n &=& \frac{1}{N}\sum_{\mu=1}^N\langle\Psi(t_{\rm f})|
  \gamma^{\dagger}_{\mu,{\rm f}}\gamma_{\mu,{\rm f}}|
  \Psi(t_{\rm f})\rangle,  \nonumber\\
 Q &=& \langle\Psi(t_{\rm f})| H_F(t_{\rm f}) |\Psi(t_{\rm f})\rangle
  - E_g. \label{nqeq1}
\end{eqnarray}
Next, we switch to the Fermion operators $\alpha$ from $\gamma_{\rm
f}$ using Eq.\ (\ref{gammaf})and shift to the Heisenberg
representation. Substituting the expansion Eq.\ (\ref{BdG1}) in Eq.\
(\ref{nqeq1}), one obtains
\begin{eqnarray}
 n &=& \frac{1}{N}{\rm tr} \left[
 \left(v^{T}_{\rm f}u(t_{\rm f})+u^{T}_{\rm f}v(t_{\rm f})\right)
 \left(v^{\dagger}(t_{\rm f})u^{\ast}_{\rm f}
  + u^{\dagger}(t_{\rm f})v^{\ast}_{\rm f}\right)
\right], \nonumber\\
 Q &=& 2\sum_{\mu=1}^N\epsilon_{\mu,{\rm f}} \nonumber\\
 && \times \left[
 \left(v^{T}_{\rm f}u(t_{\rm f})+u^{T}_{\rm f}v(t_{\rm f})\right)
 \left(v^{\dagger}(t_{\rm f})u^{\ast}_{\rm f}
  + u^{\dagger}(t_{\rm f})v^{\ast}_{\rm f}\right)
\right]_{\mu,\mu}. \label{nqeq2} \nonumber\\
\end{eqnarray}

First, we present numerical results for
two cases of the quench. For each value of the quench time $\tau$,
simulations were carried for $16$ different configurations of
$\{D_{\vec{n}}\}$ for obtaining a large enough sample set for
disorder averaging. Figure \ref{fig:n-Q} shows the disorder
averaged values of density of excitations and residual energy as a
function of $\tau$ after the time evolution. The results of
simulation suggest that for large $\tau$, the density of
excitation $n$ and residual energy $Q$ scale with $\tau$ as
\begin{eqnarray}
 n &\sim& \tau^{-1/2}, \quad
 Q \sim \tau^{-1},
\end{eqnarray}
for an evolution ending inside a gapless phase and
\begin{eqnarray}
 n  &\sim& \tau^{-1/2}, \quad
 Q \sim  \tau^{-1/2},
\end{eqnarray}
for that ending in a gapped phase after passing through the gapless
phase. We note that these scaling laws are different from those
obtained for uniform $D_{\vec n}$ \cite{sen1}.

\begin{figure}[tbp]
\begin{center}
\includegraphics[width=7cm,clip]{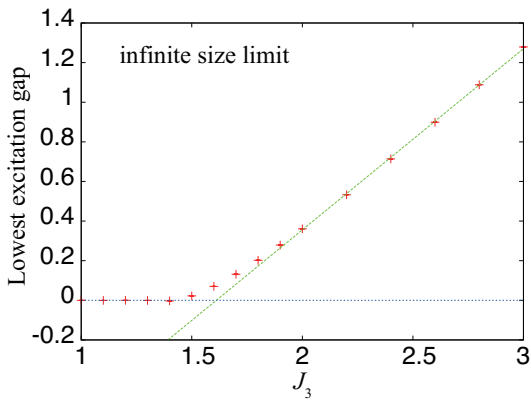}
\end{center}
\caption{The excitation gap $\Delta$ in the thermodynamic limit
estimated by the finite-size scaling. The gap vanishes for $J_3$
less than $J_3\sim 1.4$, while it increases with $J_3$ almost
linearly for $J_3$ larger than $J_3\sim 1.5$. Although some
ambiguity exists between $J_3\sim 1.4$ and $1.5$, the critical point
lies around $J_{3, c} \simeq 1.5$. Since the gap increases
 as $\Delta\propto (J_3-J_{3,c})$, one should have $z\nu = 1$.
} \label{fig:Gap-J3}
\end{figure}

\begin{figure}[tbp]
\begin{center}
\includegraphics[width=7cm,clip]{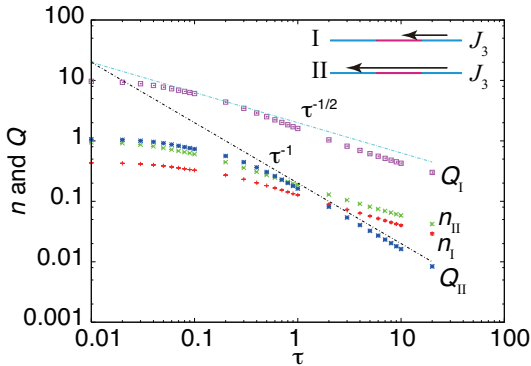}
\end{center}
\caption{Scalings of the density of excitations and residual energy
after a quench of $J_3$ from $5$ to $-5$ and from $5$ to $0$. The
density of excitations is scaled as $n_{ex}\sim\tau^{-1/2}$ in both
cases. The scaling of residual energy is $Q\sim\tau^{-1}$ when $J_3$
stops at $5$ and $Q\sim\tau^{-1/2}$ when $J_3$ stops at $0$.
Simulations are carried out for systems with $16\times 16$ unit
cells. The average is taken over 16 configurations of
$\{D_{\vec{n}}\}$. } \label{fig:n-Q}
\end{figure}

A qualitative explanation of such scaling laws for $n$ and $Q$ can
be obtained as follows. We recall that for dynamics in critical
systems without disorder, the condition for diabaticity is given by
$\frac{d\Delta}{dt}\ge \Delta^2$ (Ref. \onlinecite{pol1}). In generic
second order quantum phase transition with critical exponents $z$
and $\nu$, one can write $\Delta\sim\lambda^{z\nu}$ where $\lambda$
is quenched with a rate $1/\tau$. This yields standard expressions
\cite{pol1} $\hat{\Delta} \sim \tau^{-z\nu/(z\nu+1)}$. From this,
one can estimate the scaling form of the density of excitations and
the residual energies to be
\begin{eqnarray}
n \sim\int_0^{\hat{\Delta}} D(\varepsilon)d\varepsilon, \quad Q \sim
\int_0^{\hat{\Delta}} (\Delta_{\rm f} + \varepsilon) D(\varepsilon)
d\varepsilon,
\end{eqnarray}
where $D(\varepsilon)$ is the density of states of quasi-particles
near the critical point or gapless region and $\Delta_{\rm f}$ is the
final excitation gap when the quench stops. Note that $\Delta_f=0$
for a quench ending in the gapless region. Typically, the density of
states at the critical point or in a gapless region is given by
$D(\varepsilon) \sim \varepsilon^p$ for some non-negative exponent $p$.
Using this, one may obtain scaling of the density of excitation and
the residual energies as
\begin{eqnarray}
 n \sim \hat{\Delta}^{p+1}\sim \tau^{-(p+1)z\nu/(z\nu+1)}, \nonumber\\
 Q \sim \hat{\Delta}^{p+2} \sim \tau^{-(p+2)z\nu/(z\nu+1)},
\end{eqnarray}
where in the second line we have assumed that $\Delta_{\rm f}=0$. For
finite $\Delta_f$, $n$ and $Q$ scales according to the same power
law.

\begin{figure}[tbp]
\begin{center}
\includegraphics[width=7cm,clip]{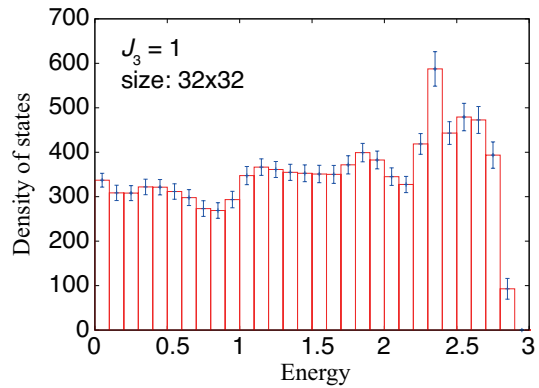}
\end{center}
\caption{Density of states of quasi-particles at $J_1 = J_2 = J_3 =
1$ in gapless phase. The quasi-particle energies $\epsilon_{\mu}$
are computed for systems with $32\times 32$ unit cells and the
histogram of them is obtained. The bin is set at $0.1$. The average
is takes over $10000$ instances of $\{D_{\vec n}\}$. The density of
states with small energy takes a finite value and fluctuates. The
amplitude of fluctuation is comparable with the error bars. This
result suggests $D(\varepsilon)$ is a constant when $\varepsilon$ is
small. } \label{fig:DOS}
\end{figure}

To obtain the scaling of the gap, we need to obtain the value of
$p$. To this end, we plot the density of states for a finite-sized
system with $32 \times 32$ unit cells in Fig. \ref{fig:DOS}. The
plot suggests that the density of states is a constant at least at
low energies $\epsilon/J \le 0.5 $ implying $p=0$ for the critical
modes. We have checked that this holds for other system sizes as
well. Moreover, numerical studies shown in Fig. \ref{fig:Gap-J3}
leads to $z\nu = 1$. Using these facts, one obtains
\begin{eqnarray}
n  \sim \tau^{-1/2}, \quad Q \sim \left\{\begin{array}{@{\,}cl}
\tau^{-1} & \mbox{gapless phase} \\
 \tau^{-1/2} & \mbox{gapped phase}.
\end{array}
\right. \label{disscal}
\end{eqnarray}

\begin{figure}[tbp]
\begin{center}
 \includegraphics[width=7cm,clip]{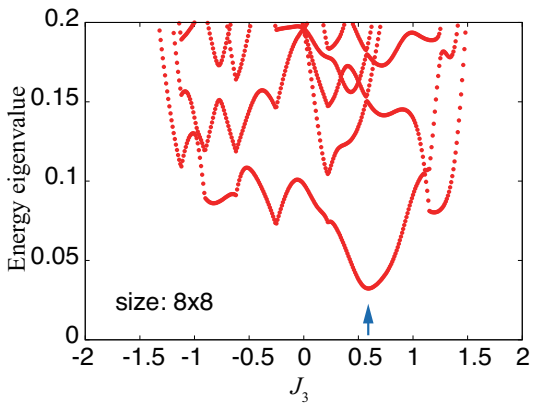}
\end{center}
\caption{Low-lying positive eigenvalues of $M$ as a function of
$J_3$ with a fixed configuration of
 $\{D_{\vec{n}}\}$. The system is composed of $8\times 8$ unit cells.
} \label{fig:Eigenvalues}
\end{figure}

The scaling laws in Eq.\ (\ref{disscal}) can be also obtained by
another argument. To elucidate this, we show, in Fig.
\ref{fig:Eigenvalues}, the low-lying energy spectra of
quasi-particles of the model as a function of $J_3$ for finite sized
system ($8\times 8$ unit cells) for a single configuration of
$\{D_{\vec{n}}\}$. Since there is a finite gap for all values of
$J_3$, an adiabatic evolution do not lead to quasi-particle
excitation. For non-adiabatic processes, the most probable
excitation occurs around the avoided level crossing with minimum
energy gap shown with a blue arrow in Fig. \ref{fig:Eigenvalues}. We
denote the corresponding energy gap by $\Delta_l$. The probability
of excitation is well approximated by the Landau-Zener formula:
$e^{-c\Delta_l^2\tau}$, where $c$ is a constant factor determined
by the slope of the excitation gap around $\Delta_l$. Next, we
recall that the distribution of excitation gaps $\Delta$ for fixed
$J_3$ inside the gapless phase scales as $1/L^2$. Hence the
distribution of $\Delta_l$ is also a function of $\Delta_l L^2$.
Thus the probability distribution function of $\Delta_l$ can be
written as $P(u)$ with $u=\Delta_l L^2$. Assuming that the factor
$c$ is independent of $L$ and $\Delta_l$, the averaged probability
of excitation $n$ is given by
\begin{eqnarray}
 n &\sim& \int_0^{\infty} du P(u)e^{-c\Delta_l^2\tau}
= \int_0^{\infty} du P(u)e^{-cu^2\tau/L^4} \nonumber\\
&=& \Pi (\tau/L^4). \label{scan}
\end{eqnarray}
From this, one can obtain a length $L_{\varepsilon}$ that yields
averaged probability of excitation $\Pi=\varepsilon$ for a given
$\tau$: $L_{\varepsilon} =
\left(\frac{\tau}{\Pi^{-1}(\varepsilon)}\right)^{1/4}$. For
sufficiently small $\varepsilon$,
$N_{\varepsilon}=L_{\varepsilon}^2$ is regarded as the average size
within which a single excitation is expected to occur. The density
of these excitations is thus estimated by $N_{\varepsilon}$ as
\begin{equation}
n  \sim \frac{1}{N_{\varepsilon}} =
\left(\frac{\Pi^{-1}(\varepsilon)}{\tau}\right)^{1/2} \propto
\tau^{-1/2}.
\end{equation}
Note that these arguments do not depend on whether the quench ends
inside the gapless phase or not since for slow dynamics the defects
are produced mostly during the passage through the gapless regime.
Using the fact that $p=0$ for these systems, a similar analysis for
$Q$ reproduces the results of Eq.\ (\ref{disscal}).

\section{ Discussion}
\label{conc1}

In conclusion, we have shown that the Kitaev model constitutes an
example of a two-dimensional model with an anisotropic critical
point. We have also demonstrated that the presence of such an
anisotropic critical point leads to novel scaling laws defect
density and residual energy during slow power-law dynamics which
takes the system from a gapped phase to the vicinity of such a
critical point. We have generalized our results for such scaling
laws for $d$-dimensional systems with such anisotropic critical
point. Further, we have computed all independent correlation
functions of the Kitaev model in the Fermionic representation after
a slow linear ramp which brings the system to the vicinity of an
anisotropic critical point. We have charted out the spatial
dependence of the correlation function and discussed its relation
with several multiple spin correlators of the model. Finally, we
have studied the non-equilibrium slow dynamics of the disordered
Kitaev model where disorder is introduced via random configuration
of $D_{\vec n}$ in its Fermionic representation. We have shown
numerically that the defect density $n$, generated during a slow
linear ramp from a gapped phase of the model to either a gapless
phase or to another gapped phase through a gapless region, scales as
$\tau^{-1/2}$. In contrast, the residual energy $Q$ scales as
$\tau^{-1/2}$ ($\tau^{-1}$) for similar dynamics ending on the
gapless surface (gapped phase after passing through the gapless
surface). We provide a qualitative understanding of such scaling
laws to back up our numerical results. We note that there has been
suggestions of experimental realization of the Kitaev model using
ultracold atomic system \cite{blochrev}. In the event of such a
realization, the simplest experimental test of our theory would
involve measurement of defect density $n$ following a slow ramp.
Such experiments has recently been performed for standard ultracold
boson systems \cite{greiner1}.

The authors thank A. Dutta, K. Kubo, A. Polkovnikov, G. Santoro and
D. Sen for discussions.  KS thanks DST, India for support through
grant SR/S2/CMP-001/2009. SS acknowledges support from Grant-in-Aid
for Scientific Research from MEXT, Japan.

\vspace{-\baselineskip}

\end{document}